\documentclass[conference]{IEEEtran}
\IEEEoverridecommandlockouts
\usepackage{cite}
\usepackage{amsmath,amssymb,amsfonts}
\usepackage{algorithmic}
\usepackage{graphicx}
\usepackage{textcomp}
\usepackage{xcolor}
\usepackage{hyperref} 
\def\BibTeX{{\rm B\kern-.05em{\sc i\kern-.025em b}\kern-.08em
    T\kern-.1667em\lower.7ex\hbox{E}\kern-.125emX}}
\begin{document}

\title{Real-time Safety Assessment of Dynamic Systems in Non-stationary Environments: A Review of Methods and Techniques}

\author{\IEEEauthorblockN{Zeyi Liu}
\IEEEauthorblockA{\textit{Department of Automation} \\
\textit{Tsinghua University}\\
Beijing, China \\
liuzy21@mails.tsinghua.edu.cn}
\and
\IEEEauthorblockN{Songqiao Hu}
\IEEEauthorblockA{\textit{School of Automation} \\
\textit{Beijing Institute of Technology}\\
Beijing, China \\
1120193091@bit.edu.cn}
\and
\IEEEauthorblockN{Xiao He}
\IEEEauthorblockA{\textit{Department of Automation} \\
\textit{Tsinghua University}\\
Beijing, China \\
hexiao@tsinghua.edu.cn
}
\thanks{This work was supported by the National Natural Science Foundation of China under grant 61733009, National Key Research and Development Program of China under grant 2022YFB25031103, and Huaneng Group Science and Technology Research Project. (\emph{Corresponding author: Xiao He.})}
}

\maketitle

\begin{abstract}
Real-time safety assessment (RTSA) of dynamic systems is a critical task that has significant implications for various fields such as industrial and transportation applications,  especially in non-stationary environments. However, the absence of a comprehensive review of real-time safety assessment methods in non-stationary environments impedes the progress and refinement of related methods. In this paper, a review of methods and techniques for RTSA tasks in non-stationary environments is provided. Specifically, the background and significance of RTSA approaches in non-stationary environments are firstly highlighted. We then present a problem description that covers the definition, classification, and main challenges. We review recent developments in related technologies such as online active learning, online semi-supervised learning, online transfer learning, and online anomaly detection. Finally, we discuss future outlooks and potential directions for further research. Our review aims to provide a comprehensive and up-to-date overview of real-time safety assessment methods in non-stationary environments, which can serve as a valuable resource for researchers and practitioners in this field.
\end{abstract}


\section{Introduction}

The dynamic system safety assessment method plays a crucial role in detecting and identifying potential safety issues in the system, providing early warnings, preventing safety incidents, and enhancing the reliability and safety of the system \cite{zio2014integrated,li2017new,PengKX-2022IJSS-Safety}. Generally, the safety level of a dynamic system can be defined as the \emph{impact degree of its operation on potential damage to surrounding people and the environment.} 
Effective safety assessment method provides early warnings to avoid safety accidents and improve the reliability and safety of the system. For instance, in the case of deep-sea manned submersibles which operate in complex deep-sea environments, the safety of such large-scale equipment not only determines the success of each dive mission but also determines the life safety of the submersible crew \cite{cui2013development,zhang2018use}. Hence, the safety assessment technology of dynamic systems is crucial to ensure the safe and reliable operation of deep-sea manned submersibles \cite{du2017safety}. Traditional safety assessment methods are classified into two categories: 1) \emph{qualitative analysis} and 2) \emph{quantitative analysis}. Qualitative analysis evaluates system safety based on some qualitative information of the system using methods such as information fusion \cite{liu2021safety,guo2021multi,feng2020new,liu2021network} and fuzzy theory \cite{zhou2020application,deng2011risk,purba2014fuzzy,liu2022measure}. Quantitative analysis, such as Petri net, Markov state transition modeling, and Bayesian network, models the safety risk based on performance data and state data obtained from monitoring.

\begin{figure*}[htbp]
  \centering
      \includegraphics[width=0.95\textwidth]{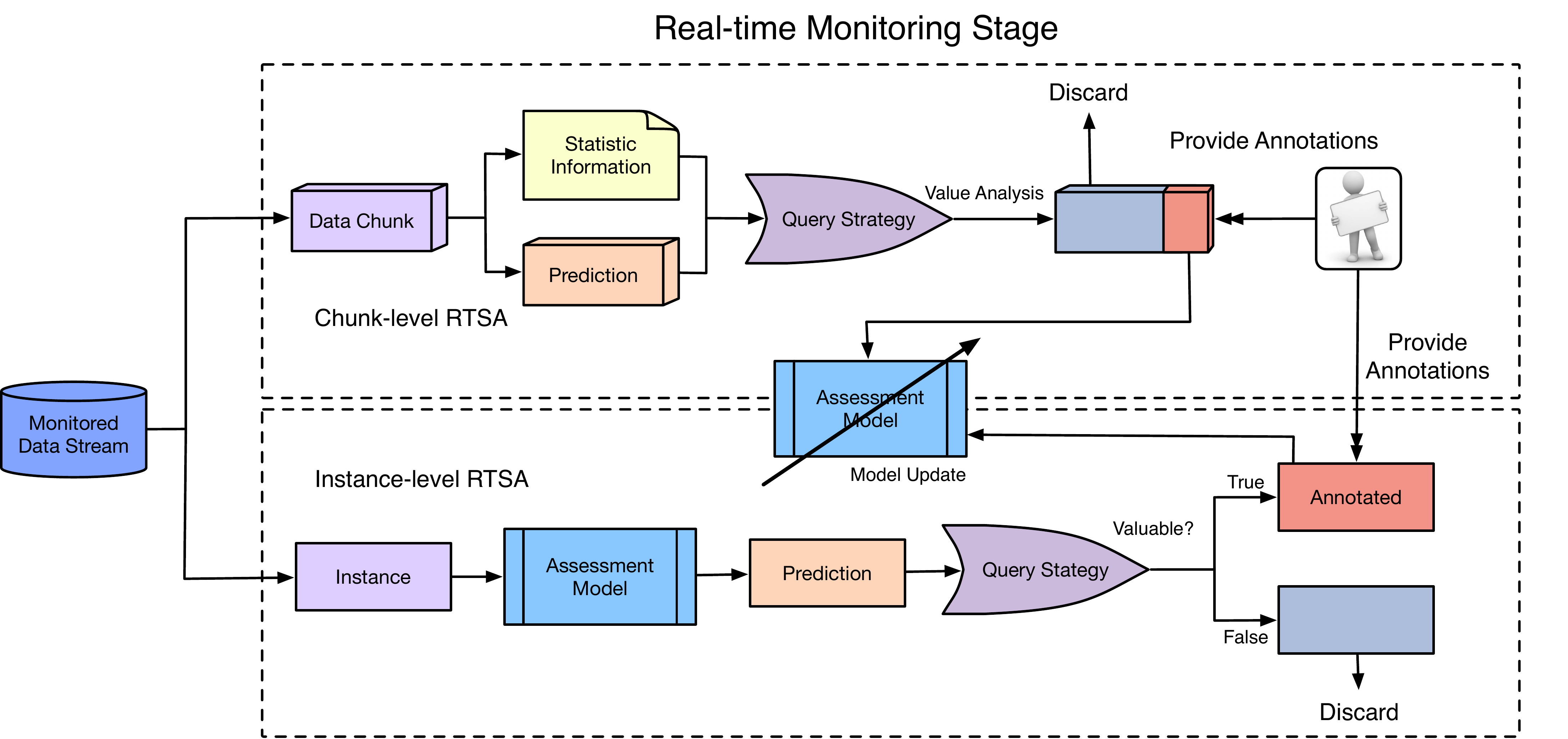}
  \caption{The typical schemes of two kinds of RTSA tasks.}\label{Fig_flowchart}
\end{figure*}

However, in non-stationary environments, traditional safety assessment methods may suffer from negative impacts \cite{ditzler2015learning}. In most practical applications, the safety of a dynamic system may not only be affected by its own internal health state but also by the environment in which it operates \cite{sayed2012learning}. For example, the non-linearity, strong coupling characteristics, multi-closed-loop control structure of deep-sea manned submersibles, and the strong uncertain disturbance of the deep-sea working environment pose new challenges to existing methods. \emph{Real-time safety assessment} (RTSA) methods are crucial to evaluating the safety of the submersible's overall system in real-time, particularly for catastrophic failures, to guarantee the life safety of the submersible crew through the dump self-rescue system. Despite the success of traditional safety assessment methods in many fields, the development of real-time safety assessment methods for non-stationary environments is still in a relatively preliminary stage. Many problems brought about by real-time safety assessment methods for non-stationary environments also need to be resolved urgently \cite{khandekar2020non,zhou2022open,parmar2023open}. Thus, exploring how to better apply existing technologies to relevant problems is still an open and valuable direction to explore.

This review provides an overview of the current state of the art in real-time safety assessment of dynamic systems in non-stationary environments. The focus is on identifying the challenges and limitations of existing methods and techniques, and exploring potential solutions and areas for improvement. The review also highlights the importance of incorporating advanced technologies and approaches, such as online learning approaches \cite{hoi2021online}, to enhance the accuracy and efficiency of real-time safety assessment.

\section{Problem Description}

\subsection{Definition}

Mon-stationary environments refer to situations where the underlying dynamics governing the system change over time, which may due to various reasons, such as changes in the system parameters, external disturbances, or sudden events. In the context of industrial applications, non-stationary environments are common in manufacturing processes, where the system dynamics can change due to variations in the raw materials, machine wear and tear, and changes in the production line setup. For example, in the semiconductor manufacturing process, variations in the process parameters can lead to changes in the device characteristics, affecting the overall quality of the product. Similarly, in transportation applications, non-stationary environments can arise due to changes in the traffic patterns, road conditions, and weather conditions. For instance, sudden changes in weather conditions such as heavy rainfall or snow can affect the road conditions and the driving behavior, leading to an increased risk of accidents.


One of the key characteristics of RTSA is its real-time nature, which requires the monitoring and analysis of system behavior to be performed in a timely and efficient manner. This poses significant challenges, especially in non-stationary environments, where system behavior can change rapidly and unpredictably. As a result, RTSA methods need to be adaptive and able to handle non-stationary data streams.
Generally, RTSA involves the \emph{continuous monitoring and analysis of system behavior to detect potential safety hazards in real-time}. The real-time safety level of a dynamic system should be predicted based on the monitoring data and environment factors, which can be typically defined as \emph{the impact degree of its operation on potential damage to surrounding people and the environment.} In this case, potential analysis of risks plays an important role in related technologies. It is important to note that RTSA is not limited to any specific type of dynamic system and can be applied to a wide range of domains, including but not limited to industrial processes and transportation systems.

\subsection{Classification}

Clearly, it is crucial for the effectiveness of RTSA tasks based on the analysis of data stream. Considering the the processing capacity, specific requirements, and characteristics of the system being monitored, two types of RTSA tasks can be divided, which are:
\begin{enumerate}

	 \item[$\bullet$] \emph{Chunk-level RTSA (CRTSA)}: CRTSA methods process the data streams in batches, which allows for more efficient processing and the use of statistical techniques that require a larger sample size \cite{lu2019adaptive}. It is especially important when dealing with high-frequency data streams, where processing each individual sample would be computationally intensive.

	 \item[$\bullet$] \emph{Instance-level RTSA (IRTSA)}: IRTSA process each individual sample separately, which allows for a more detailed analysis of each data point. It is beneficial when the system behavior is highly dynamic and requires immediate detection of safety hazards. Instance-level methods may also be more suitable for low-frequency data streams, where the amount of data is smaller and processing individual samples is more feasible. Considering the granularity of information processing, IRTSA often brings higher labeling costs and computational complexity.

\end{enumerate}

The main difference between these two approaches is the amount of information that is analyzed. Chunk-level RTSA tasks are typically used when the sampling frequency is very high and a large amount of data is being processed. In this case, statistical analysis and other indicators are used to analyze the data in chunks, which allows for a more efficient analysis. In contrast, instance-level RTSA tasks focus on analyzing each individual sample in the data stream. As illustrated in Figure \ref{Fig_flowchart}, the primary logic flow of CRTSA and IRTSA differs. CRTSA primarily employs statistical information to aid in the query process, while IRTSA typically relies on prediction to assess the value of samples. While this approach involves more information than chunk-level analysis, it typically yields higher overall evaluation results.

In this context, it is important to note that the choice between these two types of RTSA tasks depends on the specific application and the nature of the data being analyzed. Careful consideration should be given to the trade-off between the amount of information analyzed and the accuracy of the evaluation results.

\subsection{Main Challenges}

Although the considerable importance of RTSA in non-stationary environments, several prominent issues require careful consideration. These issues include, but are not limited to, the following:

\subsubsection{Constraint of Limited Annotations}

One of the main challenges for RTSA tasks in non-stationary environments. is the constraint of limited annotations. In many real-world applications, the amount of labeled data available for training is often limited, making it difficult to build accurate models. Such a limitation can result in poor model performance, especially when dealing with complex data sets or when the data distribution changes over time. Namely, the feedback of humans is essential but limited.

\subsubsection{Model Design with Incremental Update}

Another challenge is designing models with incremental update capabilities. RTSA approaches requires models that can be updated with new data in real-time, while still maintaining their accuracy and consistency. Hence, it requires careful consideration of the model architecture, training algorithm, and data representation to ensure that the model can adapt to newly arrived data.

\subsubsection{Existence of Concept Drifts}

The existence of concept drifts is another challenge for RTSA tasks in non-stationary environments. Concept drifts occur when the statistical properties of the data change over time, making it difficult to maintain model accuracy \cite{lu2018learning,vzliobaite2016overview,li2023evidential}. Such a challenge requires the development of methods that can detect and adapt to concept drifts, such as incorporating drift detection and retraining strategies into the learning process \cite{2023CADM}.

\subsubsection{Data Quality and Characters}

The quality and characteristics of the data can pose a challenge. In some cases, the data stream may be noisy, high-dimensional or incomplete, making it difficult to extract meaningful patterns. In other cases, the data stream may also be highly imbalanced or non-stationary, requiring specialized techniques for handling these types of data. Addressing these challenges requires careful consideration of data pre-processing techniques and the selection of appropriate learning frameworks\footnote{Given the benefits that datasets offer for researchers to test the performance of their algorithms and develop more accurate and reliable safety assessment systems, we have made two datasets available as open source. These datasets include simulation datasets with concept drift and realistic operating process datasets for dynamic systems in non-stationary environments. For more details, please refer to \href{https://github.com/THUFDD/JiaolongDSMS_datasets}{https://github.com/THUFDD/JiaolongDSMS\_datasets} and \href{https://github.com/THUFDD/THU-Concept-Drift-Datasets}{https://github.com/THUFDD/THU-Concept-Drift-Datasets}.}.


\section{Related Technologies}

\subsection{Online Active Learning}

Online active learning is an essential technique that facilitates real-time safety assessment by intelligently selecting and querying the most informative data samples. It reduces the need for manual annotation while maintaining the learning efficiency of the model \cite{9869794}. With the aid of online active learning, real-time safety assessment models can adaptively incorporate new data, effectively improving their accuracy and responsiveness to changing environments \cite{lughofer2017line,liu2022active}.

In IRTSA tasks within non-stationary environments, the online active learning approach enables the evaluation of individual data points as they are received in real-time \cite{liu2022online}. Consequently, this assessment method may prompt annotation requests and model update processes during critical moments, such as when drift occurs. The model can then undergo incremental updates based on a select few valuable samples. Given the dynamics of the underlying distribution and the presence of noise interference, such a paradigm may, in some cases, yield superior results in comparison to supervised learning techniques \cite{liu2023TITS}.

Regarding chunk-level real-time safety assessment tasks in non-stationary environments, two strategies can be mainly considered. On the one hand, we can approximate the necessity of labeling incoming data chunks by assessing their value. Entire data chunks may be discarded when deemed unnecessary. On the other hand, we can allocate annotations to high-value samples at a fixed proportion within each incoming block, utilizing pool-based active learning approaches in such cases \cite{settles2009active}. As a result, the primary challenge lies in evaluating the value of incoming data chunks and the samples they contain. When compared to IRTSA, such a paradigm can more effectively leverage statistical information to facilitate the design of value estimators.


In recent years, reinforcement learning has increasingly been applied to online active learning and may become a prominent development trend in the coming years. Balancing exploration and exploitation is particularly suitable for scenarios with concept drift. While blind exploitation can consistently yield satisfactory historical results, as concepts shift, those once satisfactory results may no longer be adequate. Exploration, however, effectively addresses this issue, continuously uncovering new information and potentially discovering improved solutions when concept drift occurs \cite{bouneffouf2014contextual,kim2022active,kim2020date}.




\subsection{Online Semi-supervised Learning}

In the context of real-time safety assessment, online semi-supervised learning serves as an effective approach that leverages both labeled and unlabeled data. By efficiently utilizing the abundant supply of unlabeled data available, this method allows models to generalize better, enhancing their performance on safety-related tasks. Furthermore, online semi-supervised learning enables continuous model improvement, reducing the time and effort required for manual data labeling \cite{liao2023novel}.

In general, \textit{semi-supervised learning} (SSL) constitutes a significant class of machine learning techniques that seek to leverage unlabeled data for learning tasks. This approach has been extensively researched in batch learning settings, with detailed surveys available \cite{van2020survey}. However, in the context of online learning, there are two main areas of research. The first is the transformation of traditional batch SSL methods into online algorithms capable of processing data streams comprising labeled and unlabeled data. We refer to this as \emph{online manifold regularization}. The second area of research explores classical online learning tasks in transductive learning settings, where the assumption is that unlabeled data is available before online learning tasks commence. This area is called \emph{transductive online learning}. 

It is worth noting that online active learning is a specialized form of online semi-supervised learning, wherein an online learner handles both labeled and unlabeled data. Several studies integrate these two approaches into a unified framework.



\subsection{Online Transfer Learning}

Online transfer learning is another vital technology that bolsters real-time safety assessment capabilities. By leveraging pre-trained models and knowledge acquired from related tasks, online transfer learning accelerates the learning process and helps safety assessment models adapt to new environments quickly. This rapid adaptation is particularly crucial in safety-critical applications where timely and accurate decision-making is of paramount importance.

In order to enhance the performance of models in target domains, two prevalent methods are employed: sample weighting and feature matching \cite{long2014transfer}. Sample weighting involves reusing samples from the source domain, evaluating their values, assigning corresponding weights, and then training new models or updating existing ones accordingly. In recent research, Jin \textit{et al.}  trained weighted networks through source-target joint meta-learning, assigning higher weights to source regions that aid in target fine-tuning \cite{jin2022selective}. A double-weighted multi-source transfer learning approach , termed DRMTL, was then proposed based on a regularized least squares classifier \cite{ji2019novel}. Ahmadvand \textit{et al.} applied a novel instance re-weighting scheme based on graph optimization to match the value distribution of source sample weights \cite{ahmadvand2021metric}. While in \cite{li2023multi}, a kernel mean matching method founded on individual correlation was introduced for sample weighting.

Feature matching, on the other hand, addresses the issue of differing feature spaces across domains by mapping the source domain to the target domain, or mapping both domains to a new one. The objective is to ensure the marginal and conditional distributions of source and target domains are as consistent as possible while preserving properties and identifying correlations between features. To measure feature similarity between source and target domains, metrics must be introduced. Commonly used indicators include \textit{Maximum Mean Discrepancy} (MMD), \textit{Kullback-Leibler Divergence}, and \textit{Jensen-Shannon Divergence} \cite{arbel2019maximum, yang2021learning}. Given the exceptional feature extraction capabilities of neural networks, they are highly suited for this method \cite{long2018transferable, zhong2019novel, liang2020transfer}.



\subsection{Online Anomaly Detection}

Finally, online anomaly detection plays a pivotal role in real-time safety assessment by monitoring data streams and identifying irregularities or unexpected behaviors that may signify safety hazards. By providing early warnings and facilitating proactive interventions, this technique ensures safety by allowing for timely corrective actions to be taken in response to potential threats.
In a non-stationary environment, the occurrence of a change results in an anomaly in comparison to the environment before the change. The use of online anomaly detection technology to detect changes in the environment and promptly adjust the model can enhance decision-making by providing more accurate recommendations \cite{saurav2018online}.

Online anomaly detection technology can be broadly categorized into two types: sample-based and distribution-based methods. The sample-based methods involve identifying abnormal samples from a set of samples, whereas distribution-based methods detect changes in the distribution of a set of samples relative to another set of samples. In the former technique, it is common to measure the degree of deviation of a sample from the mean. For instance, the Otsu algorithm, which is used for image binarization, employed the maximum variance of black and white pixels to establish the threshold for identifying abnormal samples \cite{lu2018outlier}. Hundman \textit{et al.} utilized an index that maximizes the mean, standard deviation, and continuous sequence to determine the threshold and perform anomaly detection \cite{hundman2018detecting}. Siffer \textit{et al.} used the extreme value theory to detect anomalies, fit the distribution of extreme values using available data, and identify new data as abnormal based on the distribution \cite{siffer2017anomaly}. Additionally, the Gaussian distribution is often used in sample-based anomaly detection techniques \cite{dwivedi2021gaussian, wu2022fl, liang2019quantum}.
Distribution-based techniques, on the other hand, usually utilize statistical methods to determine changes in the distribution of samples. Hypothesis testing is a popular statistical method used to assess whether two distributions are equivalent \cite{list2019multiple}. The relative entropy methods are commonly used to measure the difference between two distributions \cite{wang2011statistical, kullback1951information}.



\section{Future Outlooks and Potential Directions}

The future of RTSA approaches lies in developing more robust and adaptive online learning algorithms that can address the challenges such as limited annotations, incremental update, concept drifts, and data quality and characteristics. In particular, there is a need for the development of algorithms that can handle complex and heterogeneous data streams, such as multi-modal and multi-task data. In this section, several potential directions of the future research regarding RTSA tasks in non-stationary environment are summarized as follows:
\
\begin{enumerate}

\item \emph{Model interpretability}: The interpretability of online learning models has always been a very challenging problem in actual industrial scenarios. With the development of information technology, \emph{eXplainable artificial intelligence} (XAI) technique has gained widespread attention, which aims to explain how the learning model understands the data in the decision-making process. Interpretable models can increase trust in the decision-making process, improve model performance, and facilitate the discovery of new insights. Therefore, model interpretability methods have high research value, especially in dynamic or online scenarios.

\item \emph{Distributed data streams}: With the increasing amount and velocity of data generated by various sources, it is essential to have scalable algorithms that can handle the high volume and velocity of data streams. Additionally, the distributed nature of many complex dynamic systems requires the ability to process and analyze data across multiple nodes and clusters, which further emphasizes the need for scalable and efficient algorithms. The development of such frameworks can improve the accuracy and efficiency of safety assessment, allowing for quicker and more informed decision-making in response to potential safety hazards.

\item \emph{Misbehavior analysis}: For the complex dynamic systems with human-in-the-loop characteristics, misbehavior is unavoidable, which may cause real-world problems such as mislabeling to negatively affect model updates. Regarding the problem of noise in annotations, future research can focus on developing methods that can automatically identify and correct noisy annotations. Once potential misbehavior is detected, the model should provide early warning and ask experts to make relevant auxiliary judgments, so as to improve the accuracy of RTSA results.

\item \emph{Verification delay}: In actual tasks, since the data stream is continuously collected, there may be a certain time delay in the process of improving the model with annotations provided by humans. In the future, the development of advanced algorithms and techniques that can process data streams in real-time with minimal delay should be explored, which could involve the use of faster processors, more efficient algorithms, or the development of novel data compression and filtering techniques. Another approach could be to investigate the use of edge computing architectures to reduce latency and processing times.

\end{enumerate}

\section{Conclusion}

In conclusion, we have provided a comprehensive review of RTSA methods and techniques for dynamic systems in non-stationary environments. Our analysis has highlighted the importance and challenges of RTSA in various fields, such as industrial and transportation applications, and has reviewed recent developments in related technologies. We hope that this review will have served as a valuable resource for researchers and practitioners, enabling them to develop more effective and efficient approaches for RTSA tasks. We also anticipate that future research in this area will continue to focus on improving the accuracy, reliability, and adaptability of RTSA methods, making them more suitable for non-stationary environments.

\bibliographystyle{ieeetr}

\end{document}